# Evolutionary Optimization in Code-Based Test Compression

Ilia Polian    Alejandro Czutro    Bernd Becker

Albert-Ludwigs-University
Georges-Köhler-Allee 51
79110 Freiburg i. Br., Germany
{polian|aczutro|becker}@informatik.uni-freiburg.de

## Abstract

*We provide a general formulation for the code-based test compression problem with fixed-length input blocks and propose a solution approach based on Evolutionary Algorithms. In contrast to existing code-based methods, we allow unspecified values in matching vectors, which allows encoding of arbitrary test sets using a relatively small number of codewords. Experimental results for both stuck-at and path delay fault test sets for ISCAS circuits demonstrate an improvement compared to existing techniques.*

## 1  Introduction

Test data compression becomes increasingly popular for distributing test complexity between automatic test equipment and on-chip structures. By transferring compressed information from the tester and decompressing it on-chip, the requirements on IC's pin throughput can be significantly reduced while on-chip hardware cost is kept in check. Most test compression techniques proposed so far are based either on codes or on reseeding.

*Code-based* approaches assign a codeword $C(v)$ to a sequence of test data $v$ (an *input block*). For instance, $v$ could be 10011101, and $C(v)$ could be 110. Then, if 110 is transmitted from the tester, an on-chip decoder would generate the sequence 10011101 and apply it to the circuit under test. Schemes based on run-length codes [1], selective Huffman codes [2], Golomb codes [3], frequency directed codes [4], VIHC codes [5], LZ77 [6], mutation codes [7], packet-based codes [8] and non-linear combinational codes [9] have been proposed.

*Reseeding* is a technique that is applied to a pseudo-random pattern generator in order to omit those parts of its sequence that are not needed for testing. If the control information (on which patterns to omit) is fed by the tester, reseeding becomes a test compression method. Pseudo-random pattern generators used in reseeding architectures so far include LFSRs [10, 11], multiple-polynomial LFSRs [12, 13], twisted-ring counters [14] and folding counters [15].

Further test compression techniques include *Embedded Deterministic Test* [16], which is based on a ring generator unfolding low-bandwidth test data from the tester and feeding them into the scan chains of the circuit under test. It has been proposed to use a reconfigurable switch [17] or a reconfigurable interconnection network [18] for this purpose. A test compression architecture based on reconfiguring some of the scan flip flops into pseudo-random pattern generators is presented in [19].

In general, reseeding-based solutions embed test vectors into a pseudo-random sequence, which makes them attractive for testing combinational circuits. However, their applicability to sequential (non-scan), delay or functional test sets is limited. Embedded deterministic test is also designed for full scan-based circuits. In contrast, code-based compression techniques precisely reproduce the original encoded test set (possibly specifying the don't-care bits); they do not reorder the test set or add new vectors to it. Hence, code-based methods have applications for which they are inherently better suited than other compression techniques, irrespective of the achieved compression rates.

Recently, Tehranipour et al. [20] have proposed a fixed-length input block compression technique based on exactly nine codes (*9C compression*). The codes are defined with respect to an even number $K$ which is the length of an input block. For instance, let $K$ be 6. Then, the test set is partitioned into 6-bit long pieces, i.e. input blocks. Every input block is mapped to one of the following nine *matching vectors* (MVs): $v^{(1)} = 000\ 000$; $v^{(2)} = 111\ 111$; $v^{(3)} = 000\ 111$; $v^{(4)} = 111\ 000$; $v^{(5)} = 111\ UUU$; $v^{(6)} = UUU\ 111$; $v^{(7)} = 000\ UUU$; $v^{(8)} = UUU\ 000$; $v^{(9)} = UUU\ UUU$. Every vector $v^{(i)}$ is assigned a codeword $C(v^{(i)})$. For example, if 111000 appears in the test sequence, it will be encoded by $C(v^{(4)})$.

The symbol $U$ denotes an unspecified value. For instance, 111100 and 111011 both match $v^{(5)}$ (as do other 6 input blocks). In order to distinguish between these 8 possible input blocks, the fill values for the $U$ positions are spec-ified explicitly after the codeword. Thus, the input block 111100 will be coded as $C(v^{(5)})100$, and 111011 will be coded as $C(v^{(5)})011$. It is possible to represent 111000 both as $C(v^{(4)})$ and as $C(v^{(5)})000$, although the first code will probably be shorter. It is also possible to encode 111000 as $C(v^{(8)})111$ or as $C(v^{(9)})111000$. Note that any arbitrary input block can be encoded using $v^{(9)}$. However, the number of $U$s in an MV corresponds to the number of fill values that have to be applied after the codeword for that vector; hence, it is better to use MVs with as few $U$ values as possible.

In this paper, we generalize the use of the unspecified value $U$ in fixed-length input block code-based test compression. We allow MVs with 1, 0 and $U$ values on arbitrary positions instead of the fixed MV set $v^{(1)}, \ldots, v^{(9)}$ used in 9C compression. We also do not require the number of MVs (and thus codewords) to be 9. So, if the only input blocks starting with 111 are 111100 and 111110, we can use $1111U0$ instead of $111UUU$, saving two fill values for each such code. Furthermore, if the test set string includes many input blocks 110100 and 110000, an efficient set of MVs should contain the matching vector $110U00$ (in 9C compression, MVs $UUUUUU$ and $UUU000$ would be required, respectively). In general, our technique is advantageous when there are input blocks that "almost" match.

Given a test set (that can contain don't-care values), we are looking for a number $L$ of MVs of length $K$ that ensure best compression. $K$ and $L$ are inputs of the algorithm. Since any of the test vectors can have a 0, a 1 or a $U$ on any of its $K$ positions, the space of all possible alternatives is huge. Hence, we employ *Evolutionary Algorithms* (EAs) [21, 22] for optimization. EAs have been employed in many areas of VLSI CAD and design automation [23, 24], including test pattern generation [25, 26] and BIST configuration [27, 28], but not in test compression. We assign codewords to the determined MVs using Huffman coding [29]. Experimental results for ISCAS circuits show a substantial improvement compared to 9C compression.

The remainder of the paper is organized as follows: the next section formally defines the fixed-length input block compression with unspecified values. The compression method, including EA-based calculation of matching vectors, is discussed in Section 3. Experimental results are reported in Section 4. Section 5 concludes the paper.

## 2 Problem Formulation

Suppose that the circuit has $n$ inputs and the number of test patterns is $T$. Let $tp^{(i)}$ denote the $i^\text{th}$ test pattern, and let $tp_j^{(i)}$ denote the $j^\text{th}$ bit position of the $i^\text{th}$ test pattern ($tp_j^{(i)} \in \{0, 1, X\}$; $1 \leq i \leq T$; $1 \leq j \leq n$). Note that $X$ denotes a don't-care value that can be set to 0 or 1 without violating the fault coverage targets. The overall number of bit positions in a test set is $T \cdot n$. We consider the complete test set as one large string of information $t_1 t_2 \ldots t_{T \cdot n} := tp_1^{(1)} tp_2^{(1)} \ldots tp_n^{(2)} \ldots tp_1^{(T)} \ldots tp_n^{(T)}$. Our target is to efficiently encode this string of information.

Let $K$ be a positive integer number. (Note that in contrast to 9C compression it is not required to be odd.) The test set string $t_1 \ldots t_{Tn}$ is partitioned into fixed-length *input blocks* of length $K$, as formalized in the next definition. $K$ should be a divisor of the length $Tn$ of the test set string; if that is not the case, the test set string is filled up by adding $(K \cdot \lceil Tn/K \rceil - Tn)$ $X$ values in the end.

**Definition**: An *input block* is a subsequence $t_{(j-1)K+1} t_{(j-1)K+2} \ldots t_{jK}$ of the test set string $t_1 \ldots t_{Tn}$ (which has the length $K$), where $1 \leq j \leq Tn/K$ (remember that $K$ is a divisor of $Tn$). The $k^\text{th}$ input block is denoted as $ib^{(k)} \in \{0, 1, X\}^K$. □

The fixed-length input block encoding is based on *matching vectors* to which the input blocks are mapped if they *match*.

**Definition**: A *matching vector* (MV) $v$ consists of $K$ bit positions $v_1 v_2 \ldots v_K$, where $v_j \in \{0, 1, U\}$, $1 \leq j \leq K$ (where $U$ stands for "unspecified"). An MV $v$ *matches* an input block $ib$ if there is no bit position $j$ for which either $ib_j = 1$ and $v_j = 0$ or $ib_j = 0$ and $v_j = 1$. (I.e. 1 matches with 1, 0 matches with 0, and $X$ and $U$ match with arbitrary values). □

Let $L$ be the number of MVs, and let the set of MVs be $v^{(1)}, v^{(2)}, \ldots, v^{(L)}$. Note that as the MVs can contain the $U$ value, different MVs can match the same input block.

Every MV $v^{(i)}$ is assigned a *codeword* $C(v^{(i)}) \in \{0, 1\}^*$. We allow the codewords to have different length, but we require the whole code $\{C(v^{(1)}), C(v^{(2)}), \ldots, C(v^{(L)})\}$ to be a prefix code, i.e. no codeword is identical to a beginning of a different codeword. An input block that matches an MV is encoded by the codeword of that MV followed by the values in the input block corresponding to unspecified values ($U$s) in the MV:

**Definition**: Let a matching vector $v$ have $N_U(v)$ unspecified values $U$ at the positions $u_1, \ldots, u_{N_U(v)}$ (i.e. $v_{u_1} = \ldots = v_{u_{N_U(v)}} = U$), $0 \leq N_U(v) \leq K$. Let an input block $ib$ be matched by $v$. The *encoding* $C(ib, v)$ of $ib$ by $v$ is given by $C(v) ib_{u_1} ib_{u_2} \ldots ib_{u_{N_U(v)}}$. The *length of the encoding* $|C(ib, v)|$ is the sum of the length of $C(v)$ and the number $N_U(v)$ of the unspecified values in $v$. Note that $|C(ib, v)|$ is independent from $ib$. □

Our goal is to efficiently encode a test set string $t_1 t_2 \ldots t_{Tn}$ (which aggregates the whole test set) divided into $Tn/K$ input blocks. We formulate the following subproblems:

**Problem Formulation**  Given the test set (i.e. the sequence of input blocks $ib_1, ib_2, \ldots, ib_{Tn/K}$), the size $K$ and the number of MVs $L$, determine:

1. $L$ MVs $v^{(1)}, v^{(2)}, \ldots, v^{(L)}$, where $v^{(i)} \in \{0, 1, U\}^K$, $1 \leq i \leq L$, and

2. a prefix code $\{C(v^{(1)}), \ldots, C(v^{(L)})\}$,

such that the sum of the encoding lengths of the input blocks is minimal. □

Note that the problem solved by Tehranipour et al. [20] is a special case of our generic formulation: they use $L = 9$, a fixed set of MVs (specified in the introduction) and also a fixed encoding for the MVs.

## 3 Solution Approach

We solve the problem formulated above by iteratively applying the following process to the sequence of input blocks corresponding to the test set:

1. Determine $L$ matching vectors (MVs).

2. **(Covering)** Assign an MV to each input block; calculate the frequency-of-use for each MV.

3. **(Encoding)** Encode the input blocks using the collected frequency-of-use data.

Suppose that the $L$ MVs are fixed. Covering is done by selecting an input block and identifying an MV that matches it. If multiple MVs match an input block, we select the one with the lowest number of $U$s, as the encoding is more compact for such MVs (Note that it is possible that none of the $L$ MVs matches the input block; in this case, encoding is impossible with this set of $L$ MVs. This can be ruled out by setting all positions of one of the MVs to $U$.) Encoding is done using Huffman's algorithm.

The solution space for determining $L$ MVs of dimension $K$ (i.e. the number of different sets of $L$ $K$-dimensional vectors over $\{0, 1, U\}$) is $3^{KL}$ (or $3^{KL}/(L!)$, observing that the ordering of the test vectors is irrelevant), which is a huge number even for relatively small values of $K$ and $L$. Hence, we apply an evolutionary algorithm for this sub-problem. Details are reported in the sections below.

### 3.1 Matching vector determination

We employ Evolutionary Optimization [21, 22] to determine $L$ matching vectors (MVs). We define an *individual* as a set of $L$ $K$-dimensional MVs, i.e. a string $v_1^{(1)} v_2^{(1)} \ldots v_K^{(1)} v_1^{(2)} \ldots v_1^{(L)} \ldots v_K^{(L)}$ of length $KL$ over the alphabet $\{0, 1, U\}$. Figure 1 shows the pseudocode for an Evolutionary Algorithm (EA). The algorithm starts with a random *initial population* of $S$ individuals. For each individual, its *fitness* is calculated by running covering and encoding procedures (described below) and determining the reached compression rate, which is given as $100\% \cdot$ (original test set size $-$ size of compressed data) / (original test set size). Higher fitness corresponds to a higher encoding efficiency. Fitness of an individual for which covering is impossible is

```
Generate random population (S individuals);
for each individual i ∈ population
    f(i) := compression rate achieved by matching
            vectors corresponding to i's genes;
repeat {
    Generate C children, using evolutionary operators;
    for each child c
        f(c) := compression rate achieved by matching
                vectors corresponding to c's genes;
        New population := S individuals with best fitness;
}
until (termination condition fulfilled);
return individual with best fitness;
```

Figure 1: Evolutionary Algorithm

set to a sufficiently small number, such that it is lower than the fitness of an individual leading to a valid solution.

$C$ new individuals (*children*) are generated by applying *evolutionary operators* (namely crossover, mutation and inversion) to randomly selected individuals from the initial population. The fitness of each of the $C$ children is determined in the same way as for the individuals in the initial population. From these $S + C$ individuals, $S$ having highest fitness (thus leading to the highest compression) are selected to form the new population (second generation).

The *crossover* operator takes two individuals (parents) and produces two children by exchanging bit positions (*genes*) of the parents. Note that for our encoding, an individual has $KL$ genes. The resulting children have genes of one parent in several positions and the genes of the other parent in others. The *mutation* operator generates one child from one parent by replacing one randomly selected gene of a parent by a random value. The *inversion* operator produces a child by reverting the *ordering* of the genes between two random positions of a parent.

The evolutionary optimization process stops when a *termination condition* is satisfied. We used limits on the number of generated legal solutions and on the number of generations in which no fitness improvement was registered as termination conditions. Then, the MV set given by the fittest individual is returned.

### 3.2 Covering

Once the $L$ matching vectors (MVs) are generated, they are sorted in the order of increasing number of $U$s: $N_U(v^{(1)}) \leq N_U(v^{(2)}) \leq \cdots \leq N_U(v^{(L)})$. When matching an input block $ib$, this sorted list is processed, and the first matching MV $v^{(i)}$ is taken. We will comment on the implications of this technique on the optimality of the overall algorithm in the next section. For every MV $v^{(i)}$, we calculate its *fre-* *quency* $F_i$, i.e. the number of input blocks encoded by $v^{(i)}$.

### 3.3 Encoding

Given the $L$ MVs $v^{(1)}$, ..., $v^{(L)}$ and their frequencies $F_1, \ldots F_L$, the optimal prefix code is obtained by Huffman coding [29]. However, the problem formulation is slightly different here: first, an MV with a frequency of 0 can be simply left out without allocating a codeword to it. Second, due to the existence of the unspecified values ($U$s) one MV may be subsumed by another. The following example demonstrates a situation in which the Huffman coding algorithm leads to a suboptimal solution due to these specifics:

**Example**: Suppose that the MVs are $v^{(1)} = 111U$ with frequency $F_1 = 5$, $v^{(2)} = 1110$ with frequency 3 and $v^{(3)} = 0000$ with frequency 2. Huffman coding yields $C(v^{(1)}) = '0'$, $C(v^{(2)}) = '10'$ and $C(v^{(3)}) = '11'$. Since $v^{(1)}$ contains one $U$ value, it is encoded by $|'0'| + 1 = 2$ bits; $v^{(2)}$ and $v^{(3)}$ are encoded by $|'10'| = 2$ bits and $|'11'| = 2$ bits, respectively, resulting in the total size of compressed data of $5 \cdot 2 + 3 \cdot 2 + 2 \cdot 2 = 20$ bits.

However, 3 input blocks that are matched by $v^{(2)} = 1110$ are also matched by $v^{(1)} = 111U$. If we omit $v^{(2)}$ altogether, we can encode $v^{(1)}$ as $'0'$ and $v^{(3)}$ as $'1'$. Since $v^{(1)}$ now subsumes $v^{(2)}$, its frequency is now $5+3 = 8$. Its encoding is still 2 bits long, while the encoding of $v^{(3)}$ is 1 bit long. The total size of compressed data is $8 \cdot 2 + 2 \cdot 1 = 18$ bits, which is less than 20 bits calculated by the Huffman algorithm. □

The suboptimal result of the example is a consequence of the covering algorithm from the previous section: if the three input blocks had been covered by $v^{(1)}$ rather than $v^{(2)}$, the Huffman algorithm would have produced the optimal result. Handling such cases explicitly could improve the compression rate.

## 4 Experimental Results

We applied our method to uncompacted stuck-at test sets with don't-cares obtained by the method from [30] for ISCAS 85 and combinational parts of ISCAS 89 circuits, and to uncompacted path delay test sets with don't-cares for combinational parts of ISCAS-89 circuits generated by the tool TIP [31, 32] (with 100% robust path delay coverage). We used the package GAME [33] for evolutionary optimization.

Table 1 summarizes the results for stuck-at test sets. The first two columns contain the name of the circuit and the size of the test set (note that the circuits are sorted with respect to increasing test set size). All other columns quote the compression rate $100\% \cdot \frac{\text{original test set size} - \text{size of compressed data}}{\text{original test set size}}$ of various methods. The compression is better for higher values of this number.

| Circuit | Test set size | Compression rate | | | |
|---|---|---|---|---|---|
| | | 9C | 9C+HC | EA | EA-Best |
| s349 | 624 | 23.0% | 30.0% | 54.2% | **55.8%** |
| s344 | 624 | 25.0% | 33.0% | 51.8% | **55.8%** |
| s298 | 629 | 19.0% | 27.0% | 45.2% | **51.2%** |
| s208 | 722 | 26.0% | 32.0% | 47.8% | **50.4%** |
| s400 | 984 | 29.0% | 36.0% | 54.4% | **56.4%** |
| s382 | 1008 | 29.0% | 36.0% | 52.0% | **54.2%** |
| s386 | 1157 | 0.0% | 13.0% | 30.4% | **30.6%** |
| s444 | 1176 | 40.0% | 43.0% | 54.4% | **57.8%** |
| c6288 | 1216 | 8.0% | 19.0% | 17.6% | **20.4%** |
| s510 | 1850 | 42.0% | 45.0% | **57.6%** | **57.6%** |
| c432 | 1944 | 26.0% | 36.0% | 49.2% | **50.4%** |
| s526 | 1944 | 25.0% | 29.0% | **46.4%** | **46.4%** |
| s1494 | 2324 | -1.0% | 11.0% | 23.0% | **28.9%** |
| s420 | 2380 | 53.0% | 55.0% | 54.4% | **56.2%** |
| s1488 | 2436 | 2.0% | 15.0% | 25.6% | **30.0%** |
| s832 | 3404 | 35.0% | 38.0% | **43.8%** | **43.8%** |
| s820 | 3496 | 31.0% | 35.0% | 42.8% | **43.4%** |
| c499 | 3854 | 43.0% | 51.0% | 45.0% | **51.6%** |
| s713 | 4104 | 51.0% | 52.0% | 61.4% | **61.8%** |
| s641 | 4212 | 51.0% | 52.0% | 60.2% | **62.2%** |
| c880 | 4680 | 40.0% | 42.0% | 47.8% | **49.8%** |
| c1908 | 4950 | -2.0% | 10.0% | 18.4% | **19.0%** |
| s953 | 5220 | 51.0% | 53.0% | 61.6% | **63.2%** |
| c1355 | 5289 | 38.0% | **45.0%** | 40.8% | 44.8% |
| s1196 | 6016 | 34.0% | 38.0% | **46.2%** | **46.2%** |
| s1238 | 6240 | 34.0% | 37.0% | 44.0% | **45.8%** |
| s1423 | 8463 | 59.0% | 59.0% | 61.0% | **61.6%** |
| s838 | 8509 | 67.0% | 68.0% | 66.2% | **68.6%** |
| c3540 | 10350 | 36.0% | 39.0% | 43.8% | **44.2%** |
| c2670 | 33086 | 70.0% | 70.0% | 70.4% | **70.6%** |
| c5315 | 33108 | 65.0% | 65.0% | 66.2% | **67.0%** |
| c7552 | 60030 | 63.0% | **64.0%** | 63.2% | 63.2% |
| s5378 | 71262 | 73.0% | 73.0% | **76.8%** | **76.8%** |
| s9234 | 118560 | 75.0% | 75.0% | 76.2% | **76.4%** |
| s35932 | 133988 | 71.0% | 71.0% | **73.8%** | **73.8%** |
| s15850 | 305500 | 80.0% | 80.0% | **83.0%** | **83.0%** |
| s13207 | 410200 | 83.0% | 83.0% | 85.8% | **85.9%** |
| s38584 | 1250256 | 82.0% | 82.0% | **86.2%** | **86.2%** |
| s38417 | 2068352 | 84.0% | 84.0% | 87.0% | **87.9%** |
| Average | | 42.6% | 46.8% | 54.2% | 55.9% |

Table 1: Experimental results for stuck-at test sets

Column '9C' reports the compression rate obtained by our reimplementation of the 9C compression method from [20] for $K = 8$ (which yielded best results in that paper). Note that our numbers do not match the figures in that paper, as we use different test sets. Since the compression rates are higher for the test sets that we use than the rates reported in [20], it would be unfair to quote those results. We did not implement further compression methods, however [20] contains comparisons with several techniques.

| Circuit | Test set size | Compression rate | | | |
|---|---|---|---|---|---|
| | | 9C | 9C+HC | EA1 | EA2 |
| s27 | 448 | -5.0% | 9.0% | 46.2% | **51.6%** |
| s298 | 6018 | 41.0% | 44.0% | 48.9% | **54.2%** |
| s386 | 6032 | 8.0% | 19.0% | 24.7% | **26.0%** |
| s208 | 7524 | 40.0% | 43.0% | 43.5% | **46.6%** |
| s444 | 14544 | 49.0% | 52.0% | 55.6% | **55.8%** |
| s382 | 16272 | 50.0% | 55.0% | 58.0% | **59.2%** |
| s400 | 16320 | 50.0% | 55.0% | 57.1% | **58.2%** |
| s526 | 17088 | 44.0% | 45.0% | 59.3% | **60.0%** |
| s349 | 17712 | 41.0% | 44.0% | 57.0% | **61.2%** |
| s344 | 17712 | 41.0% | 44.0% | 57.0% | **60.8%** |
| s510 | 18450 | 45.0% | 47.0% | 48.9% | **52.6%** |
| s1494 | 20300 | 1.0% | 15.0% | 19.9% | **25.0%** |
| s1488 | 20664 | 2.0% | 15.0% | 20.5% | **24.6%** |
| s820 | 21850 | 34.0% | 38.0% | 38.2% | **42.4%** |
| s832 | 22448 | 34.0% | 38.0% | 38.4% | **42.4%** |
| s420 | 43588 | 58.0% | **59.0%** | 57.9% | 51.2% |
| s713 | 56376 | 61.0% | 63.0% | 64.6% | **69.0%** |
| s953 | 75510 | 57.0% | 59.0% | 59.4% | **62.8%** |
| s641 | 94500 | 60.0% | 62.0% | 62.6% | **66.2%** |
| s1196 | 95616 | 40.0% | 42.0% | **46.9%** | 46.4% |
| s1238 | 96128 | 39.0% | 41.0% | **46.3%** | 45.8% |
| s838 | 269808 | **70.0%** | **70.0%** | 69.3% | 64.2% |
| s1423 | 2321592 | 49.0% | 50.0% | 51.8% | **52.8%** |
| s5378 | 3625588 | 78.0% | 78.0% | 77.5% | **81.2%** |
| s9234 | 4666324 | 81.0% | 82.0% | 80.1% | **83.2%** |
| s35932 | 7108416 | 87.0% | 87.0% | 86.7% | **91.0%** |
| s13207 | 10234000 | 85.0% | 85.0% | 85.9% | **89.6%** |
| s15850 | 36502362 | 84.0% | 84.0% | 82.7% | **86.3%** |
| s38584 | 81190512 | 87.0% | 87.0% | 67.5% | **90.0%** |
| Average | | 48.7% | 52.1% | 55.6% | **58.6%** |

Table 2: Experimental results for path delay test sets

Matching vectors (MVs) have a fixed encoding in [20]: '0' for 000000; 10 for 111111; 11000 for 000111; '11001' for 111000; '11010' for $111UUU$; '11011' for $UUU111$; '11100' for $000UUU$; '11101' for $UUU000$; and '1111' for $UUUUUU$ (for $K = 6$; the same assignment applies for other $K$ values). We ran the reimplemented 9C algorithm with this encoding replaced by Huffman coding and report the results in column '9C+HC'.

The results of our proposed approach are given in column 'EA'. We used $L = 64$ and $K = 12$; in the evolutionary algorithm, we used the population size $S$ of 10, the number of children $C$ of 5, the crossover probability of 30%, the mutation probability of 30% and the inversion probability of 10%. One of the MVs was set to all-$U$, such that there were no insolvable instances. We report the average over 5 algorithm runs.

We generated data for numerous values of $K$ and $L$. While the space does not allow us to include all data, we report our best results in the last column. Additional improvements could be achieved by varying the parameters of the evolutionary algorithm, which we did not do.

Our proposed approach outperforms the original 9C algorithm for all circuits except s838 (but our best result is higher for this circuit). This could be ruled out by adding the 9C matching vector set to the initial population (which we did not). If 9C compression is combined with Huffman coding, it outperforms our method with default values for 6 circuits, and it is more efficient than our best results for two circuits (out of 39). This can be explained by the large size of the solution space; a larger population size or a reduced evolutionary pressure might be appropriate for such circuits.

The last line of the table quotes the average values. It can be seen that our results are considerably better on average than those obtained by 9C compression, both with and without Huffman coding. It can also be seen that the difference between our technique with default values and the best compression rate is relatively small. We conclude that our algorithm is relatively stable with respect to the parameters $K$ and $L$.

The structure of Table 2 is similar to that of Table 1. We report the results for the proposed method for $L = 9$ and $K = 8$ (column 'EA1') and $L = 64$ and $K = 12$ (column 'EA2'). Both values are averages over 5 successful algorithm runs with termination after 500 populations without improvement. The All-$U$ MV was included into the population. The proposed methods yields best results for all but two circuits (s420 and s838), and the second set of parameters results in an improvement of the compression from 48.7% for the original 9C compression to 58.6%, on average.

## 5 Conclusions

We provided an Evolutionary Optimization-based solution for test set encoding with fixed-length input blocks. For the first time, our problem formulation allows unspecified values ($U$s) at arbitrary positions of the matching vectors. This enables the employment of compact on-chip decoders for arbitrary test sets. No decoder re-design is required in case of a test set modification, if an all-$U$ matching vector is used; however, the compression rate might suffer. A reconfigurable decoder, into which the codeword/matching vector information can be loaded, would solve this problem.

Experimental results show the superiority of our technique. However, further improvements are possible by fitting the parameters of the Evolutionary Optimization, such as population size and operator probabilities. Another direction for further research is the application of our method in a multiple scan chain environment.

# Acknowledgment
We would like to thank Kohei Miyase and Seiji Kajihara for providing the stuck-at test sets for ISCAS circuits containing don't-care values.